\documentclass[12pt]{iopart}
\input epsf
\begin{document}

\title[Rotational bosonic current in an optical torus]{Rotational
bosonic current in a quasi-condensate confined in an optical toroidal trap}

\author{Aranya B. Bhattacherjee$\dag$$^{*}$, E. Courtade$\dag$ and  E.
Arimondo$\dag$\footnote[4]{To whom correspondence should be
addressed (arimondo@df.unipi.it)}}

\address{$\dag $ INFM, Dipartimento di Fisica {\it E. Fermi},
Universit\`{a} di Pisa,
Via Buonarroti 2, I-56127 Pisa, Italy}

\address{$^{*}$ Department of Physics, A.R.S.D. College, University of Delhi (South
Campus), Dhaula Kuan, New Delhi-110021, India}

       \date{\today}% It is always \today, today,
                    % but any date may be explicitly specified

\begin{abstract}
We investigate the possibility of inducing a bosonic current which
is rotational ($\overrightarrow{\nabla} \times
\overrightarrow{v}\neq \overrightarrow{0}$) in a pseudo 1D
quasi-condensate confined in an optical toroidal trap. The
stability of such a current is also analyzed using hydrodynamics
approach. We find that such a current is uniform when the circular
symmetry is preserved and energetically stable when the modes of
elementary excitations are restricted to one dimension. This
scheme allows to distinguish between a quasi and a true condensate
by measuring the rotational spectra of the sample.
\end{abstract}

\maketitle

\section{Introduction}
Bose-Einstein condensates provide an important quantum system
where one can observe many important phenomena such as
superfluid-Mott insulator transition \cite{greiner} and vortices
\cite{matthews, madison1, madison2, chevy}. Recent advances in
experimental techniques \cite{gorlitz, schreck} have stimulated
interest in one-dimensional (1D) Bose gases. The 1D case leads to
some interesting physics which does not occur in higher
dimensions.  The experimental feasibility of realizing a tightly
confined (1D) trapped gas has been investigated theoretically
using Bessel light beams \cite{arlt}. It is well known that a true
BEC cannot occur at any finite temperature in an interacting
homogeneous Bose gas in 1D \cite{hohenberg}, due to long
wavelength fluctuations; nor can it occur in the limit of zero
temperature \cite{pitaevskii}, due to quantum fluctuations.
However, the presence of a trapping potential changes the density
of states at low energies, and in the weakly-interacting limit a
BEC may be formed \cite{ketterle}. Recently Petrov $\emph{et al.}$
discussed three different regimes of quantum degeneracy which can
occur in a condensate confined in 1D trap \cite{petrov}: BEC,
quasi-condensate, and Tonks-Girardeau gas. In the weakly
interacting limit a BEC exists, but as the interaction becomes
stronger the mean field energy becomes important. When this energy
becomes of the order of the energy level spacing of the trap,
fluctuations again become significant. In this regime, the system
forms a quasi-condensate, with local phase coherence, rather than
the global coherence associated with a true condensate. The
quasi-condensate density has the same smooth profile as a true
condensate but the phase fluctuates in space and time. Phase
fluctuations of the condensate are caused mainly by low energy
collective excitations \cite{petrov}. Quasi-condensates in 3D have
been observed experimentally in equilibrium \cite{dettmer,
hellweg} and non-equilibrium \cite{shvarchuck}. The fact that the
velocity of a true condensate
$\overrightarrow{v_{s}}(\overrightarrow{r},t)=\hbar/m\overrightarrow{\nabla}\phi(\overrightarrow{r},t)$
(with $\phi(\overrightarrow{r},t)$ the phase of the condensate)
leads to the important constraint namely irrotationality
$\overrightarrow{\nabla} \times
\overrightarrow{v_{s}}(\overrightarrow{r},t)=0$. On the other
hand, the normal component which is substantially higher in a
quasi-condensate as compared to a true condensate is not
constrained by the condition of irrotationality. Hence we take
advantage of this property of the quasi-condensate and propose in
this work the use of a 2D optical rotator to produce a rotational
bosonic current ($\overrightarrow{\nabla} \times
\overrightarrow{v}\neq 0$ where $\overrightarrow{v}=
\overrightarrow{v_{s}}+\overrightarrow{v_{n}}$ and
$\overrightarrow{v_{n}}$ is the velocity of the normal component)
in a quasi condensate and study the conditions under which this
current is stable. We show that using this scheme, it is possible
to distinguish between quasi-condensate and true condensate. It is
to be noted that such a current is not vortex. Such a kind of a
rotator has been proposed for cold atoms based on the principle of
transfer of angular momentum of the photons to the atoms
\cite{wright} using a TEM$_{01}$ circularly polarized
Laguerre-Gaussian beam. It has been experimentally demonstrated
that a blue-detuned hollow Bessel laser beam formed by an axicon
can be successfully used to trap laser cooled metastable xenon
atoms \cite{kulin}. Experimentally, cesium atoms have been trapped
in an all blue-detuned stack of optical rings by interference of a
standard gaussian beam with a counter-propagating hollow beam
generated by a pair of axicon \cite{depret, verkerk}. We propose
the use of such an experimental scheme with a TEM$_{01}$ beam in
order to realize the 2D optical rotator.

\section{Description of the optical torus}
\label{Description of the optical torus}

\begin{figure}[h]
\centering\begin{center}\mbox{\epsfxsize 2.0 in
\epsfbox{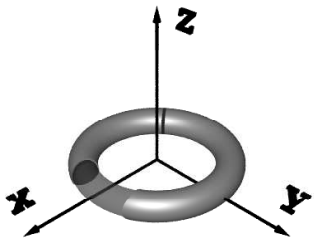}}
\end{center}
\caption{Schematic representation of the optical torus described
by Eq. (\ref{eq1}). This optical torus can be generated by the
interference of a standard gaussian beam with a
counter-propagating hollow beam generated by a pair of axicon
\cite{depret,verkerk}.} \label{torus}
\end{figure}

We will examine the motion of a condensate within an optical
toroidal ring trap as experimentally realized in Ref.
\cite{verkerk} and previously theoretically analyzed by Wright et
al. \cite{wright} The potential created by the optical laser may
be written as

\begin{equation}
V(r)=\frac{1}{2}m\omega_{r}^{2}(r-r_{0})^{2}+\frac{1}{2}m\omega_{z}^{2}z^{2}
\label{eq1}
\end{equation}

where $x^{2}+y^{2}=r^{2}$ and $r_0$ is the location of the potential
minima and also the mean radius of the torus.  $\omega_{z}$
($\omega_{r}$) is the oscillation frequency along the $z$($r$)
direction.  Because we suppose the plane of the ring perpendicular to
the direction of gravity ($z$ direction in our system), the atomic
energies are not effected by gravity.  The optical torus described by
the potential of Eq.  (\ref{eq1}) is represented in Fig.  \ref{torus}.

The atomic motion within the
toroidal trap includes also a rotational energy along the center of
the torus, described by the following rotational Hamiltonian:

\begin{equation}
H_{r}=-\frac{\hbar^{2}}{2I}\frac{\delta^{2}}{\delta \theta^{2}}
\label{eq2}
\end{equation}

where the moment of inertia for an atom with mass $m$ is given by
$I=mr_{0}^{2}$. If $p$ is the rotational quantum number, then the
rotational hamiltonian defines an energy spectrum whose energy
spacing is linked to the rotation frequency $\Omega$ for $\Delta
p=1$ as

\begin{equation}
\Omega=\frac{(2p+1)\hbar}{2mr_{0}^{2}}
\label{eq3}
\end{equation}

Note that the rotational frequency $\Omega$ depends only on the
radius of the ring $r_{0}$.  In the radial direction, the ring
radius is approximately the radius of the laser beam, which
therefore should be chosen as small as possible for a tight
confinement.  Extending the experimental parameters of
\cite{depret} to Rb atoms, with a beam radius of about 50 $\mu$m,
potential height of 10 $E_{R}$ and $\omega_{R}\sim 2\pi \times
0.34$ kHz, one can find $\omega_{r}\sim 2\pi \times 0.06$ kHz,
$\omega_{z}\sim 2\pi \times 6.7$ kHz and $\Omega\sim 2\pi \times
0.024$ Hz (for $p=0\rightarrow p=1$ transition).

\bigskip

\par
Moreover we will suppose, as in Ref. \cite{wright} that using for
the laser field an high order circularly polarized
Laguerre-Gaussian mode, an orbital angular momentum is transfered
by the laser field to the atoms.  Therefore the rotational motion
of the condensate atoms at the frequency $\Omega$ can be excited
by the laser field. Let us now systematically investigate the
collective modes of a true condensate and that of a
quasi-condensate (in the presence of a rotational velocity field)
in an optical torus.

\section{Collective modes of a true condensate in an optical torus}
\label{Collective modes of a true condensate in an optical torus}

For a true condensate the velocity is the gradient of a scalar,
the velocity field is thus irrotational, unless the phase of the
order parameter has a singularity (vortex). A simple description
of the collective oscillations is provided by the irrotational
hydrodynamics equations of superfluids

\begin{equation}
\frac{\partial n_{s}}{\partial t}+\overrightarrow{\nabla} \cdot
[n\overrightarrow{v_{s}}]=0 \label{eq5}
\end{equation}

\begin{equation}
m\frac{\partial\overrightarrow{v_{s}}}{\partial
t}+\overrightarrow{\nabla}[\frac{mv_{s}^{2}}{2}+V(\overrightarrow{r},t)+n_{s}g_{3d}]=0
\label{eq6}
\end{equation}

where $n_{s}(\overrightarrow{r},t)$ is the superfluid spatial
density, $v_{s}(\overrightarrow{r},t)$ is the superfluid velocity
field in the laboratory frame and $m$ is the mass of the atom. The
parameter $g_{3d}$ characterizes the strength of the inter-atomic
interactions and is related to the $s$-wave scattering length
$a_{s}$ for a 3D gas by $g_{3d}=4\pi\hbar^{2}a_{s}/m$. These
equations are valid in the Thomas-Fermi limit, where the so-called
quantum pressure term is neglected. The density can be written as
$n_{s}=n_{0}+\delta n_{s}$, and
$\overrightarrow{v_{s}}=\overrightarrow{v_{0}}+\delta
\overrightarrow{v_{s}}$ where $n_{0}$($\overrightarrow{v_{0}}$) is
the equilibrium density (velocity) and $\delta n_{s}$($\delta
\overrightarrow{v_{s}}$) is a small perturbation of the density
(velocity) from its equilibrium value. The collective oscillations
can be derived by looking for a general time-dependent solutions
of the form (see \cite{menotti,cozzini})

\begin{equation}
\delta
n_{s}(\overrightarrow{r},t)=\alpha_{0}+\alpha_{1}(x^{2}+y^{2})+\alpha_{2}z^{2}
+\alpha_{3}xy+\alpha_{4}xz+\alpha_{5}yz \label{eq7}
\end{equation}

\begin{equation}
\delta
\overrightarrow{v_{s}}(\overrightarrow{r},t)=\overrightarrow{\nabla}
(\beta_{0}+\beta_{1}(x^{2}+y^{2})+\beta_{2}z^{2}
+\beta_{3}xy+\beta_{4}xz+\beta_{5}yz) \label{eq8}
\end{equation}

where $\alpha_{i}$ and $\beta_{i}$ are time dependents parameters
to be determined. In the linear limit, one can look for solutions
varying in time like $e^{-i\omega t}$. Linearizing Eqs.
(\ref{eq5}) and (\ref{eq6}) and inserting $\delta n_{s}$ and
$\delta \overrightarrow{v_{s}}$ and solving for the modes $\omega$
for $\omega_{x}=\omega_{y}=\omega_{r}$, one finds

\begin{equation}
\omega_{\pm}^{2}=\frac{3}{2}(\omega^{2}_{r}+\omega^{2}_{z})\pm
\frac{1}{2}\sqrt{9\omega^{4}_{r}+9\omega^{4}_{z}-10\omega^{2}_{r}\omega^{2}_{z}}
\label{eq9}
\end{equation}

\begin{figure}[h]
\centering\begin{center}\mbox{\epsfxsize 3.1 in
\epsfbox{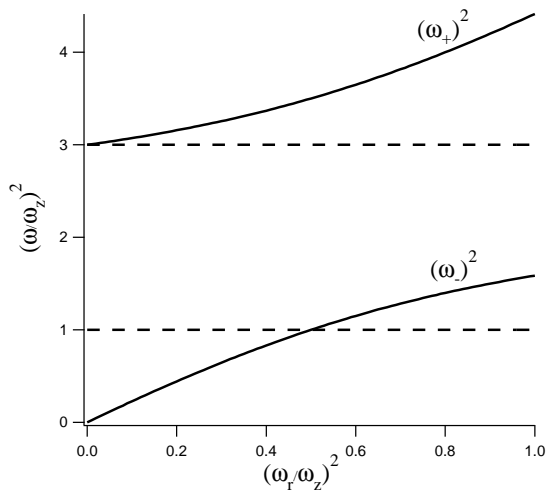}}
\end{center}
\caption{Mode frequencies of Eq. (\ref{eq9}) as functions of the
radial trap frequency $\omega_{r}$. The solid lines correspond to
the modes of the true condensate in the optical torus. The dashed
lines correspond to the modes of the condensate in the case of
tight confinement in the optical torus. The positive values of the
mode frequencies ensures stability.} \label{roots}
\end{figure}

These modes describe the coupling between the center of mass
motion along the radial and axial directions. Fig. \ref{roots}
displays these modes as a function of the radial frequency which
are seen to be stable. The frequencies are normalized with respect
to the $\omega_{z}$ axial frequency. We now allow for the
condition of tight confinement, i.e. the particles are constrained
to move only along the path $x^{2}+y^{2}=r_{0}^{2}$. In the tight
confinement regime, at sufficiently low temperatures the radial
motion of particles (the center of mass modes along $x$ and the
$y$ direction) is essentially frozen if the radial frequency
$\omega_{r}$ is much more than the mean-field interaction energy.
Note that the tight confinement regime is not the 1D regime as
discussed in \cite{petrov}. In our system, we still have the
velocity component in both $x$ and $y$ directions. We prefer to
call it as the pseudo 1D regime. Proceeding as before, we find in
the tight confinement regime the two modes corresponding to center
of mass motion along the $z$ direction and the scissoring modes
$xz$($yz$) as $\omega^{2}=3\omega^{2}_{z}$ and
$\omega^{2}=\omega^{2}_{z}$ respectively. These low energy
excitations frequencies are exactly the lowest mode of excitation
found numerically in \cite{salernich} which confirms the accuracy
of the hydrodynamic approach.

\section{Collective modes of a quasi-condensate in an optical torus}
\subsection{Rotational velocity field of a quasi-condensate}
The special properties of superfluids are a consequence of their
motions being constraint by the fact that the velocity
($\overrightarrow{v_{s}}(\overrightarrow{r},t)$) of the condensate
is proportional to the gradient of the phase
($\phi(\overrightarrow{r},t)$) of the wave function that is
$\overrightarrow{v_{s}}(\overrightarrow{r},t)=\hbar/m\overrightarrow{\nabla}\phi(\overrightarrow{r},t)$.
We see directly from this definition that
$\overrightarrow{v_{s}}(\overrightarrow{r},t)$ satisfies two
important constraints, namely the condition of irrotationality
$\overrightarrow{\nabla} \times
\overrightarrow{v_{s}}(\overrightarrow{r},t)=0$ and the
Onsager-Feynman quantization condition $\oint
\overrightarrow{v_{s}}(\overrightarrow{r},t).d\overrightarrow{l}=nh/m$
($n=0,\pm 1, \pm 2 ...$). Obviously, in a simply connected
geometry with condensate wave function finite everywhere, the
condition of irrotationality implies the Onsager-Feynman
quantization condition with $n=0$, but for more general cases
involving vortices or in a torus (multiply connected geometry),
$n$ can be non zero.

In stable or metastable equilibrium the total bosonic current will
have contribution from both the normal as well as the superfluid
component (Landau's two fluid hypothesis). Now if by some
mechanism, the condensate in the torus with mean radius $r_{0}$ is
made to rotate with an angular velocity $\Omega$, then the linear
velocity of the normal component will be simply $\Omega r_{0}$
while the linear velocity of the superfluid component will be
constrained by the quantization condition and in general can not
be equal to $\Omega r_{0}$. In fact, a simple statistical
mechanical argument shows that the lowest free energy is obtained
when $n$ takes the value closest to $\Omega/ \Omega_{c}$ where
$\Omega_{c}=\hbar/mr_{0}^{2}$. For $\Omega/ \Omega_{c}\ll1$, $n$
is equal to zero and consequently the superfluid component no
longer contributes to the circulating current. At larger values of
$\Omega$ ($>\Omega_{c}/2$), the superfluid will contribute to the
total angular momentum an amount $\sim n\Omega_{c}$. An inspection
of Eq. (\ref{eq3}) for $p=0\rightarrow p=1$ transition together
with the definition of $\Omega_{c}$, we find that
$\Omega/\Omega_{c}=0.5$ irrespective of the radius of the 2D
optical rotator. Consequently, in a true condensate, where the
normal component is extremely low, the total bosonic current is
almost zero for $\Omega/ \Omega_{c}\ll1$. Based on the above
discussion, we argue that in a quasi-condensate, where the normal
component is substantially higher than a true condensate, for
$\Omega/ \Omega_{c}\ll1$, the normal component will contribute
significantly to the bosonic current, which one could detect. Thus
using this proposal, it could be possible to distinguish between a
quasi-condensate and a true one.

We propose that the theoretical scheme introduced in \cite{wright}
may be used to transfer orbital angular momentum of the photon to
a cloud of ultra-cold atoms in a quasi-condensate state. A
velocity field created in such a manner is rotational. This
velocity field can be visualized by introducing a vector potential
$\overrightarrow{A}$ in the quasi-condensate total velocity
$\overrightarrow{v}=
\overrightarrow{v_{s}}+\overrightarrow{v_{n}}$

\begin{equation}
\overrightarrow{v}(\overrightarrow{r})=\frac{\hbar}{m}[\nabla \phi
(\overrightarrow{r})+ \overrightarrow{A}] \label{eq10}
\end{equation}

where the vector potential $\overrightarrow{A}$ is associated with
the rotational component of the velocity

\begin{equation}
\overrightarrow{A}=\frac{m}{\hbar}\overrightarrow{\Omega} \times
\overrightarrow{r} \label{eq11}
\end{equation}

with $\overrightarrow{\Omega}=\Omega \widehat{k}$. $\Omega$ is the
rotational angular velocity of the sample. The corresponding
rotational velocity field which is actually the velocity of the
normal component

\begin{equation}
\overrightarrow{v}_{rot}=\overrightarrow{v}_{n}=\Omega
(-y\widehat{i}+x\widehat{j})\label{eq12}
\end{equation}

with $\overrightarrow{\nabla} \times \overrightarrow{v}_{n}=2
\overrightarrow{\Omega}$. The angular velocity of the sample is
easily adjusted by the mean ring radius $r_{0}$ (see Eq.
(\ref{eq3})).

\subsection{Stability conditions for stable persistent current}
Let us now analyze the condition for existence of a steady
velocity field $\overrightarrow{v}_{n}$ of the form of Eq.
(\ref{eq12}) in  a quasi-condensate for a tight confinement in the
presence of a weak perturbation (a slightly distorted potential).
The perturbed potential is written as

\begin{equation}
V(r)=\frac{1}{2}m\omega_{r}^{2}(r-r_{0})^{2}+\frac{1}{2}m\omega_{z}^{2}z^{2}
+\frac{1}{2}m(\delta\omega_{x})^{2}x^{2}+\frac{1}{2}m(\delta\omega_{y})^{2}y^{2}
\label{eq13}
\end{equation}

where $\delta\omega_{x}$ and $\delta\omega_{y}$ are small external
perturbations in the frequencies along $x$ and $y$ direction. In
the presence of the perturbation, the irrotational component of
the velocity field is now also included which is written as
$\overrightarrow{v}_{irrot}=\overrightarrow{v}_{s}=\alpha
\overrightarrow{\nabla}(xy)$. Therefore the total velocity field
is

\begin{equation}
\overrightarrow{v}(\overrightarrow{r})=(\alpha -
\Omega)y\widehat{i}+(\alpha + \Omega)x\widehat{j} \label{eq14}
\end{equation}

The macroscopic description of modes with circulation is provided
by the equations of two component rotational hydrodynamics
\cite{cozzini,griffin}

\begin{equation}
m\frac{\partial\overrightarrow{v}}{\partial
t}+\overrightarrow{\nabla}[\frac{mv^{2}}{2}+V(\overrightarrow{r},t)+g_{3d}n_{s}+2g_{3d}n_{n}]=
m\overrightarrow{v} \times (\overrightarrow{\nabla} \times
\overrightarrow{v})\label{eq15}
\end{equation}

In the following, we will ignore fluctuations in the density of
the normal component ($n_{n}$) and also neglect damping due to
interaction between superfluid and normal component. The
continuity equation leads to the following expression for $\alpha$

\begin{equation}
\alpha =
\frac{\Omega((\delta\omega_{y})^{2}-(\delta\omega_{x}))^{2}}{2\Omega^{2}-((\delta\omega_{y})^{2}+(\delta\omega_{x})^{2})}
\label{eq17}
\end{equation}

Thus we see from Eq. (\ref{eq14}) that the irrotational component
of the velocity field which comes into picture as a result of the
perturbation destroys the bosonic current with uniform angular
velocity $\Omega$. From Eq. (\ref{eq17}) we infer that for such a
uniform current to exist, we must have
$(\delta\omega_{x})^{2}=(\delta\omega_{y})^{2}$. The conservation
of circular symmetry guarantees a uniform current in the
quasi-condensate. In the presence of a weak, static, asymmetric
perturbation, angular momentum is not conserved. In a rotating
quasi-condensate, scattering of particles against such an
asymmetric perturbation will ultimately bring the fluid to rest.
Having found that circular symmetry as a necessary criterion for a
uniform current to exist, we now proceed to calculate the
collective modes in the non-equilibrium state. Collective modes in
the presence of a rotational velocity field for a trap with
circular symmetry can be derived as in section \ref{Collective
modes of a true condensate in an optical torus} by looking for
time-dependant solutions for $\delta n$ and $\delta
\overrightarrow{v}$. We now allow for tight confinement and assume
that we still have motion in  $xz$($yz$) planes and also center of
mass motion along the $z$ direction. The density and velocity
fluctuations are now proportional to $z^{2}, xz, yz$. These modes
from the usual hydrodynamic approach are found to be

\begin{equation}
\omega^{2}=3\omega^{2}_{z} \label{eq18}
\end{equation}

and roots of the equation

\begin{equation}
\omega^{3}+\Omega
\omega^{2}-(\Omega^{2}+\omega^{2}_{z})\omega-\Omega(\Omega^{2}+\omega^{2}_{z})=0
\label{eq19}
\end{equation}

\begin{figure}[h]
\centering\begin{center}\mbox{\epsfxsize 3.1 in
\epsfbox{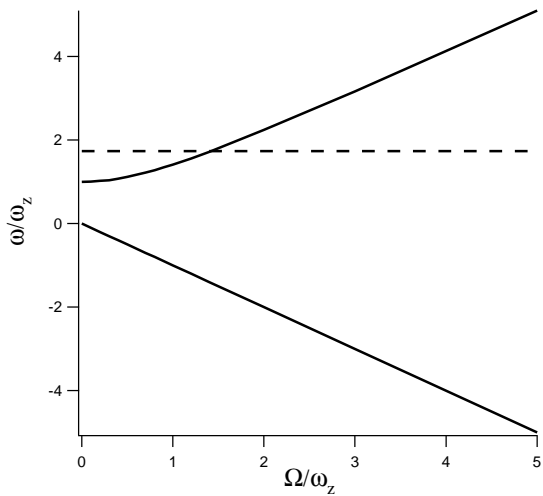}}
\end{center}
\caption{Calculated modes frequencies as functions of the angular
frequency $\Omega$. The dashed line corresponds to the center of
mass frequency along the $z$ direction from Eq. (\ref{eq18}). The
other two solid line curves correspond to the scissoring modes in
the $xz$($yz$) planes of Eq. (\ref{eq19}). The negative
frequencies of the lower curve are signature of instability (see
text).} \label{rootsbis}
\end{figure}

The modes calculated from of Eq. (\ref{eq18}) and Eq. (\ref{eq19})
are plotted in Fig. \ref{rootsbis}. Negative frequencies of
excitations are taken as signature of energetic instability which
is a result of the absence of thermodynamic equilibrium
\cite{recati}. The mode corresponding to the center of mass motion
along the $z$ direction from Eq. (\ref{eq18}) (dashed line in Fig.
\ref{rootsbis}) is always positive and constant. This non
dependence on $\Omega$ is a consequence of the absence of
excitations in the $xy$ plane due to tight confinement. We find
that for a non-zero value of $\Omega$, one of the scissoring modes
in the $xz$($yz$) planes of Eq. (\ref{eq19}) (lower curve in Fig.
\ref{rootsbis}) is always negative. This signals energetic
instability because the magnitude of the smallest root is always
equal to the angular velocity $\Omega$. This is equivalent to the
Landau criterion of super-fluid stability
\cite{javanainen1,javanainen2} with respect to proliferation of
elementary excitations i.e. the bosonic current becomes unstable
when the frequency of the lowest energy excitation becomes equal
or more than the angular velocity. A similar result is found for a
Bose gas in a toroidal container using Bogoliugov approximation
\cite{rokhsar}. Thus to have an energetically stable current we
need to freeze the motion in $xz$ and $yz$ planes. This means that
we restrict the modes of the elementary excitations to only the
$z$ direction whose frequency is given by Eq. (\ref{eq18}).

Let us now look into the possible experimental realization of such
a pseudo quasi-condensate in our proposed scheme. The mean field
interaction frequency $g_{3d}N/\hbar2\pi r_{0}^{3}$ for $N=10^{4}$
and $a_{s}=5$ nm (in the case of Rb) is found to be $\sim
2\pi\times 0.5$ Hz much less than the radial trap frequency ($\sim
2\pi\times 60$ Hz, see section \ref{Description of the optical
torus}) which implies that we are indeed in the pseudo 1D regime.
Following \cite{petrov}, one can introduce a dimensionless
quantity  $\alpha =4\pi a_{s}/l$, where
$l=\sqrt{\hbar/m\omega_{z}}$ is the amplitude of axial zero point
oscillations. The regime of a weakly interacting gas corresponds
to $\alpha\ll 1$. In this regime, the condensate is in the
Thomas-Fermi regime if the number of particles $N>\alpha^{-1}$.
The decrease of temperature to below the degeneracy temperature
$k_{B}T_{d}=N\hbar\omega_{z}$ leads to the appearance of a
quasi-condensate which at $T<T_{cr}$ turns to a true condensate.
The critical temperature $T_{cr}$ is determined by
$k_{B}T_{cr}=\hbar\omega_{z}(32N/9\alpha^{2})^{1/3}$. For $N\sim
10^{4}$ and using the values of the trap parameters given earlier
(see section \ref{Description of the optical torus}), we find
$T_{d}=3$ mK and $T_{cr}=17$ $\mu$K. Working between these two
temperature limit and having $\omega_{r}>g_{3d}N/\hbar2\pi
r_{0}^{3}$, one can thus be in the pseudo 1D quasi-condensate
regime. This scheme allows us to distinguish between a
quasi-condensate and a true condensate by simply measuring the
rotational spectrum of the sample. The rotational spectrum of the
sample can be measured by using two consecutive Bragg pulses
probing the momentum distribution in a trapped Bose gas at low
temperature \cite{stenger, brunello}. A stimulated Raman
transition involves absorption of a photon from one beam followed
by stimulated emission into the other; if the system is in a
quasi-condensate regime then this process will transfer an orbital
angular momentum off to the atom, but no linear momentum in the
axial or radial directions.

\section{Conclusion}
In conclusion, we have proposed the use of a new type of optical
toroidal trap to induce a uniform rotational bosonic current
(which is not a vortex) in a quasi-condensate. The principle of
producing such a current of rotational is based on the transfer of
angular momentum of the photon to the atoms of the normal
component. The angular velocity induced by the proposed 2D optical
rotator is exactly half the critical angular velocity required by
the superfluid component to contribute to the total bosonic
current. Hence, the given optical rotator is unable to transfer
any angular momentum to the atoms in the superfluid component. We
are thus able to distinguish between a quasi-condensate and a true
condensate by measuring the rotational spectrum of the sample by
two-photon Bragg spectroscopy. Our analysis indicates that such a
rotational current is uniform when the circular symmetry of the
ring trap is conserved and energetically stable when the
elementary excitations are completely frozen in the transverse
direction ($xy$ plane).

\section*{Acknowledgments}
Fruitful discussions with I.~Carusotto and A. Recati, and a
careful check of the manuscript by C.~Menotti are acknowledged.
This research was supported by the Abdus Salam International
Centre for Theoretical Physics, Trieste, Italy under the ICTP-TRIL
fellowship scheme and the Sezione A of INFM-Italy through a PAIS
Project, by the MIUR-Italy through a COFIN Project, and by the
European Commission through the Cold Quantum-Gases Network,
contract HPRN-CT-2000-00125.

\end{document}